# A new concept and design of ferrite-based microwave vortex devices

M. Sigalov[1], E. Kamenetskii[2], and R. Shavit[3]

*Abstract* – In microwave resonant systems with ferrite samples, one becomes faced with specific phase relations for the electromagnetic fields. Such specific phase relations may lead to appearance of nontrivial states: electromagnetic vortices. This paper provides some general ideas and design principles for microwave devices with the field vortex structures.

## 1 INTRODUCTION

General scattering and propagation characteristics of microwave waveguides containing ferrite samples were under investigations in numerous works during many years (see, e.g. [1]). Specifically, in questions of microwave theory and techniques, an interest in such systems was devoted to development of general and rigorous formulations of the problem and an analysis of properties of the scattering matrix. At the same time, the studies of quadratic relations do not touch upon a question about specific phase relations for the fields in such microwave systems.

In a case of ferrite inclusions acting in the proximity of the ferromagnetic resonance (FMR), the phase of the wave reflected from the ferrite boundary depends on the direction of the incident wave. This fact, arising from special boundary conditions for the tangential components of the fields on the dielectric-ferrite interface, causes the time-reversal symmetry breaking effect in microwave resonators with inserted ferrite samples [2]. When microwave resonators contain enclosed gyrotropic-medium samples, the electromagnetic-field eigenfunctions will be complex, even in the absent of dissipative losses. It means that the fields of eigen oscillations are not the fields of standing waves in spite of the fact that the eigen frequencies of a cavity with gyrotropic-medium samples are real [3]. This leads to very specific topological-phase characteristics. The power-flow lines of the microwave-cavity field interacting with a ferrite sample, in the proximity of its FMR, may form whirlpool-like electromagnetic vortices [4].

Vortices can be easily observed in water. They are a common occurrence in plasma science. In recent years, there has been a considerable interest in vortices of electromagnetic fields. Study of vortices with optical phase singularities has opened up a new frontier in optics [5]. In microwaves, there is a special interest in electromagnetic vortices created by ferrite particles with the FMR conditions.

It can be shown analytically that for TE polarized electromagnetic waves, the singular points of the Poynting vector (the vortex cores) are directly related to the zero-electric-field topological features in a vacuum region of the cavity space [2, 4]. To analyze the vortex structure for a case when a ferrite sample is placed in a maximal cavity electric field and to study the fields inside a ferrite region one has to use numerical simulation methods.

The main purpose of this paper is to develop novel concepts and design principles for microwave devices with the field vortex structures created by ferrite samples. In our studies, we use the HFSS (the software based on FEM method produced by ANSOFT Company) CAD simulation programs for 3D numerical modeling of Maxwell equations. In our numerical experiments, both modulus and phase of the fields are determined.

## 2 MICROWAVE VORTICES IN THE CAVITY-FERRITE-DISK SYSTEM

The most known examples of systems used for an analysis of ferrite-based microwave vortices are microwave cavities with enclosed ferrite samples [2, 4]. It was shown in [4] that in the rectangular-waveguide cavity a small ferrite disk acts as a topological defect causing induced vortices. Figure 1 gives a picture of the Poynting vector distribution in a cavity when a thin normally magnetized ferrite disk (being oriented so that its axis is perpendicular to a wide wall of a waveguide) is placed in the maximal cavity electric field. In the vicinity of a ferrite disk

Department of Electrical and Computer Engineering, Ben Gurion University of the Negev, Beer Sheva 84105, Israel,
[1] e-mail: sigalov@ee.bgu.ac.il, tel.: +972 8 6472402, fax: +972 8 6472402,
[2] e-mail: kmntsk@ee.bgu.ac.il, tel.: +972 8 6472407, fax: +972 8 6472949,
[3] e-mail: rshavit@ee.bgu.ac.il, tel.: +972 8 6471508, fax: +972 8 6472949.

and inside it, one has strong localization of the power flow density due to the vortex creation. As it is shown in Figure 1, the power input is at the left-hand side of a system. If one interchanges the microwave source and receiver positions, leaving fixed a direction of the bias magnetic field, the vortex will have the same rotation direction. So the vortex rotation direction is invariant with respect to mirror reflection along a waveguide axis. When one reverses the DC magnetic field together with an interchange of the microwave source and receiver positions, the vortex changes its rotation direction. It means that the vortex rotation direction is not invariant with respect to a combined symmetry operation: mirror reflection and time reversal. This is a distinctive feature regarding the known reciprocity-theorem relationships for gyrotropic media [6].

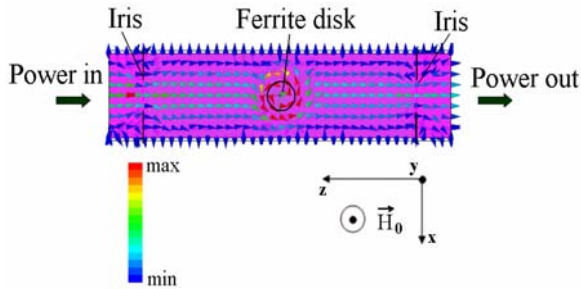

Figure 1: A rectangular-waveguide cavity with two irises and an enclosed ferrite disk.

The observed Poyning-vector vortices are characterized by unique symmetry properties of the fields. The symmetry breaking of the field structures become evident from the pictures shown in Figures 2 and 3. Figure 2 shows the RF magnetic field and Figure 3 gives the RF electric field in the vicinity of a ferrite disk. The pictures in Figures 2 and 3 have the $90°$ phase shift.

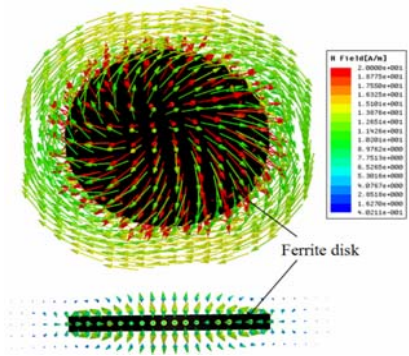

Figure 2: Top and side views of the RF magnetic field near a ferrite disk.

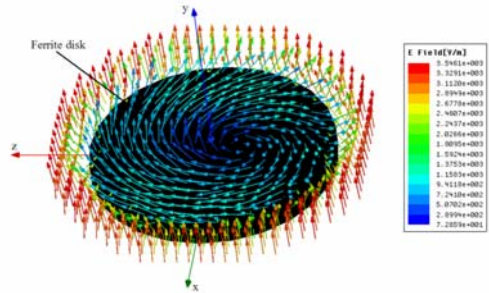

Figure 3: A view of the RF electric field near a ferrite disk.

The field localization takes place also when a ferrite disk is placed not in a maximal cavity electric field, but in a maximal cavity magnetic field. In this case, however, a structure of the Poynting-vector vortices and symmetry properties of the fields are completely different from those shown in Figures 1 – 3 [7]. It is necessary to note that in all cases the vortex does not accumulate energy. Due to the vortex one just has redistribution of energy inside a cavity.

An analysis of the Poynting-vector structure as well as the field structures is an important tool in developing design principles for novel microwave-vortex devices.

## 3 DESIGN OF MICROWAVE-VORTEX DEVICES

In attempts to use unique properties of microwave vortices for design of new microwave devices, one becomes faced with serious difficulties. The vortices are very sensitive to the field structure in a microwave system. When one strongly decreases the standing-wave ratio in a cavity, the vortex disappears. At the same time, creation of vortices in standard devices may completely destroy the functional ability of a system. Such an "anti-design" can be clearly demonstrated in a standard circulator. When one creates a vortex in a ferrite disk in a circulator, the entire system becomes completely reciprocal.

In this paper, we show some examples of proper application of the vortex concept in the design of microwave antennas and near-field microwave sensors.

### 3.1 A vortex concept in the microwave antenna design

We analyze a microwave patch antenna with an enclosed normally magnetized ferrite disk. Excitation of an antenna is due to a microstrip line. A general structure of an antenna is shown in Figure 4.

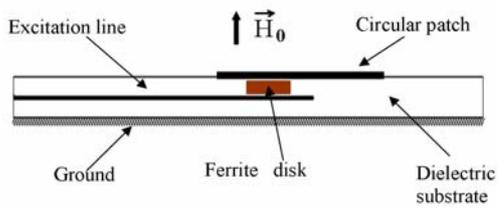

Figure 4: A microwave patch antenna with an enclosed ferrite disk.

Our main purpose is to realize a circularly polarized antenna in the best way. Figures 5 (a) and 6 (a) show the Poynting-vector pictures for two different positions of a ferrite disk. One can clearly see that the vortex structure in the second case is much better than in the first case. It is very important to see that the "vortex quality" is strongly correlated with the circular characteristics of the antenna far fields.

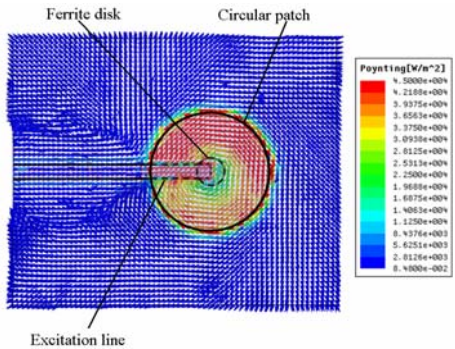

(a)

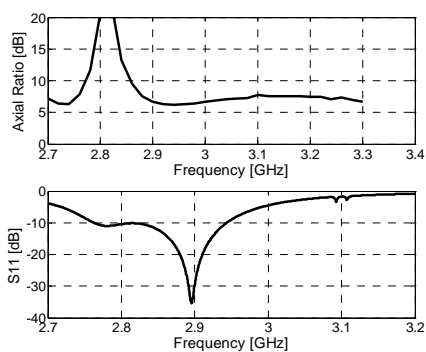

(b)

Figure 5: A ferrite disk placed at the center of a patch. (a) The Poynting-vector picture; (b) The axial ratio and $S_{11}$ parameters.

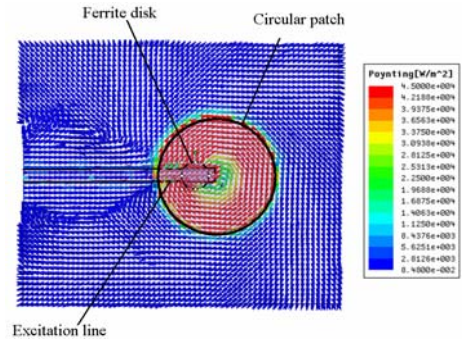

(a)

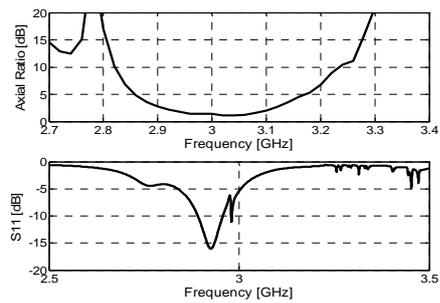

(b)

Figure 6: A ferrite disk shifted from the center of a patch. (a) The Poynting-vector picture; (b) The axial ratio and $S_{11}$ parameters.

For a "better vortex", one has better antenna properties characterizing by a very small axial ratio (see Figure 6). At the same time, when a ferrite disk is placed at the center of a patch (see Figure 5) an excitation line is better matched.

**3.2 Near-field microwave sensors**

At present, there is a strong interest in different microwave sensors for the near-field characterization of material properties (see, e.g. [8]). In design of such sensors, the vortex concept could be especially useful. The fact that due to a vortex structure one has the microwave energy localization will be very important for increasing the sensor sensitivity. On the other hand, special structures of the vortex electric and magnetic fields can be very useful for characterization of biological materials with the symmetry breaking properties. Figure 7 shows an example of a vortex-type near-field microwave sensor realized as a microstrip resonator with a ferrite cone. When the Poynting-vector vortex appears, one has strong field localization in the vicinity of a ferrite sample (see Figure 8). Figure 9 shows the Poynting-vector structure inside a ferrite cone.

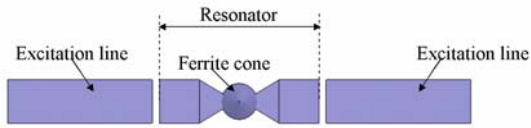

Figure 7: A vortex-type near-field microwave sensor

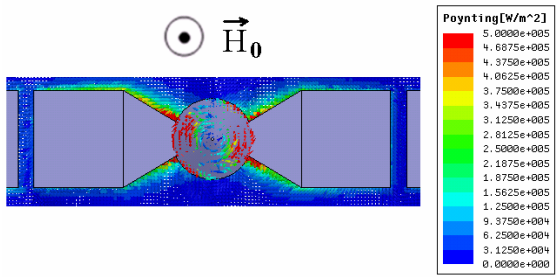

Figure 8: Field localization due to creation of the Poynting-vector vortex.

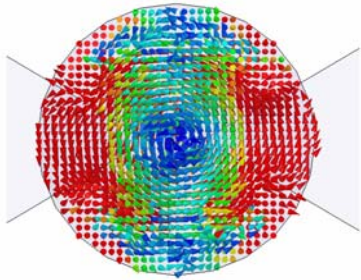

Figure 9: The Poynting-vector distribution inside a ferrite cone.

The most attractive and useful feature of this device is the rotating electric field on the tip of a cone. (see Figure 10). A direction of rotation is correlated with a direction of a bias magnetic field.

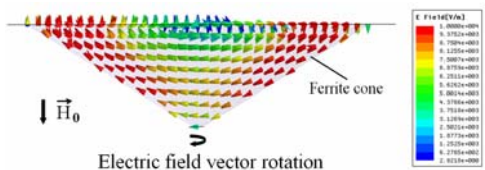

Figure 10: Rotating electric field in a ferrite cone.

## 4 CONCLUSION

In this paper we discuss a new concept and design principles for microwave devices with the field vortex structures. It becomes evident that the problem of microwave vortices of the Poynting vector created by ferrite samples can be very important for many modern applications, e.g., for processing of guided signals, for near-field microwave lenses, for field concentration in patterned microwave metamaterials, for new microwave antennas. It is supposed that these nontrivial states – the "swirling" entities – can, in principle, be used to carry data and point to new communication systems. Application of such generic ideas to microwave systems is of increasing importance in numerous utilizations.